# Automatic Ultrasound Image Segmentation of Supraclavicular Nerve Using Dilated U-Net Deep Learning Architecture

Mizuki Miyatake
*Biomedical Engineering*
*University of Florida*
Gainesville,United States
Miyatake.Mizuki.MM@gmail.com

Subhash Nerella
*Biomedical Engineering*
*University of Florida*
Gainesville,United States
subhashnerella@ufl.edu

David Simpson
*Department of Anesthesiology*
*University of Florida*
Gainesville,United States
simp214@hotmail.com

Natalia Pawlowicz
*US Anesthesia Partners*
Coppell ,United States
Natalia.Pawlowicz@usap.com

Sarah Stern
*College of Medicine*
*University of Florida*
Gainesville,United States
sstern@ufl.edu

Patrick Tighe
*Department of Anesthesiology*
*University of Florida*
Gainesville, United States
ptighe@ufl.edu

Parisa Rashidi
*Biomedical Engineering*
*University of Florida*
Gainesville, United States
parisa.rashidi@ufl.edu

*Abstract*— Automated object recognition in medical images can facilitate medical diagnosis and treatment. In this paper, we automatically segmented supraclavicular nerves in ultrasound images to assist in injecting peripheral nerve blocks. Nerve blocks are generally used for pain treatment after surgery, where ultrasound guidance is used to inject local anesthetics next to target nerves. This treatment blocks the transmission of pain signals to the brain, which can help improve the rate of recovery from surgery and significantly decrease the requirement for postoperative opioids. However, Ultrasound Guided Regional Anesthesia (UGRA) requires anesthesiologists to visually recognize the actual nerve position in the ultrasound images. This is a complex task given the myriad visual presentations of nerves in ultrasound images, and their visual similarity to many neighboring tissues. In this study, we used an automated nerve detection system for the UGRA Nerve Block treatment. The system can recognize the position of the nerve in ultrasound images using Deep Learning techniques. We developed a model to capture features of nerves by training two deep neural networks with skip connections: two extended U-Net architectures with and without dilated convolutions. This solution could potentially lead to an improved blockade of targeted nerves in regional anesthesia.

## I. Introduction

Nerve blocks are generally used for pain treatment in surgery, in which an injection of a local anesthetic is delivered around select nerves in order to block pain transmission from the site of tissue injury to the brain [1]. Prior to ultrasound, nerves were targeted based upon expert understanding of anatomy and anatomical landmarks; the needle was serially passed along a range of trajectories until it happened to pass close enough to the nerve to induce a paresthesia. Unfortunately, many of the most important nerves targeted for regional anesthesia also run in close proximity to other vital structures such as arteries, veins, lungs, and bowel, structures which may not safely accommodate one or multiple misadventures of a nerve-seeking needle.

Anesthesiologists recently use Ultrasound guidance to inject local anesthetics to target nerves. Ultrasound guidance makes it possible to differentiate nerves from adjacent tissues, visualize the needle during its transit to the desired location, and the deposition of local anesthesia around the target nerve. The advent of Ultrasound Guided Regional Anesthesia (UGRA) facilitated the use of regional anesthesia for myriad procedures in the US and other countries [2]. Moreover, ultrasound has led to the identification of several new UGRA procedures to access nerves and tissue planes that were previously considered to be challenging to access in widespread clinical practice. The rapid increase in successful programmatic UGRA services has led to accreditation of regional anesthesia and acute pain medicine fellowship programs in the United States [3].

Despite rapid increase in usage, evidence is mixed regarding the superiority of safety of UGRA over traditional landmark-based techniques to peripheral nerve block [4]. Postulated reasons include the relatively low incidence of severe complications of regional anesthetics, associated learning curve of UGRA, and residual difficulties in visualizing and tracking critical structures on US images during UGRA. Computer-assisted visualization of critical structures on ultrasound images during UGRA may improve the safety of UGRA and allow for further improvements in perioperative pain management and reduction in opioid requirements following surgery.

In this paper, we investigated the possibility of deep learning for assisting the UGRA process. Recent Deep Learning techniques achieved substantial success in general object recognition tasks [5-7] using large datasets such as

ImageNet [8], VOC Challenge [9], and SUN Database [10]. Labeled medical images, however, are typically limited in dataset size due to cost of data acquisition and annotation. Therefore, development of the neural network architecture for medical images, which does not require a large amount of data, is in high demand. We modified U-Net architecture proposed by Ronneberger et al [11] which is a CNN (Convolutional Neural Network) based architecture to predict the location of target nerve in the ultrasound images. U-Net architecture is developed to make strong use of data augmentation of available annotations.

There are many Convolutional Neural Network (CNN) based architectures widely used in image segmentation task for natural images. Segmentation in medical images requires the context of the surrounding anatomic structures to predict the target object. Yu et al. [12] developed a CNN module using dilated convolutions which support exponential expansion of receptive field without increasing the number of layers of the neural network. Increased receptive field provides context of surrounding anatomic structures. Ronneberger et al [11] proposed skip connections (U-Net) to increase the resolution of output segmentation. Skip connections provide shortcut to gradient flow in shallow networks [13]. In U-Net skip connections are used to concatenate features from contracting path to expanding path to recover the lost spatial information with the use of down sampling. Pooling and stridden convolution also increase the receptive field, but resolution is reduced. Liu. C proposes an adversarial network for Brachial plexus segmentation of upper limb[14] using dilated convolutions and data augmentation to overcome the limitations of ignored long range anatomical dependencies and elastic deformations respectively.

Inspired by prior studies, we modified U-Net architecture to include dilated convolution, bilinear interpolation along with skip connection for better resolution of the output. We also employed short skip connection similar to Resnet [15]. Short skip connections increase the convergence speed and allows to train deep neural networks [13]. Our main objective in this paper is to perform automated anatomical segmentation for UGRA, especially for the supraclavicular Brachial Plexus nerve. We compared models that recognizes target nerves automatically from ultrasound images using two different network architectures; U-Net; dilated U-Net (Dilated convolutions are introduced to U-Net. These models could potentially improve the safety and accuracy of UGRA. Our contributions to this paper are as follows: (1) we developed the first annotated dataset of Supraclavicular nerve ultrasound images. (2) We developed a modified U-Net architecture mentioned above. (3) This is the first study utilizing automated segmentation for Supraclavicular nerve detection.

## II. METHODS

### A. Data

Ultrasound videos of Supraclavicular nerve anatomical regions were obtained by using Sonosite Linear HFL50 15-6 MHz probe (Fuji Industries). The ultrasound videos of the plexus were acquired from 6 healthy subjects and were converted from mp4 format into successive images, resulting in 1118 images.

The images were manually annotated by four trained physicians, where the number of annotated images from each patient are shown in Table I. For the ease of the annotation, a built-in annotation tool of ImageJ [16] was used. The original image size was around 500x374 pixels and later resized to 128x128 pixels (Fig. 1). Annotations indicated the region which would likely be targeted for deposition of local anesthetic solution supraclavicular nerve block. The brachial plexus consists of a complex network of interwoven nerves situated between the neck and shoulder. Nerve blocks may be performed at numerous locations along the brachial plexus, including the supraclavicular region. Anatomic dissections have identified numerous and common variations in the anatomy of the brachial plexus, including the region of the supraclavicular nerve block [17]. In 61% of patients, there is asymmetry between the right and left brachial plexus within the same individual [17]. Clinically, local anesthesia is deposited in regions surrounding the brachial plexus, including proximate regions whereby the brachial plexus is normally located based upon anatomical study, but for which visual identification of nerves is less than certain. Biomedical image datasets are typically small datasets due to the time required for annotating the image. To increase the size of our dataset, we use data augmentation. We augmented the dataset via random rotations, random height and width shifts of the original images.

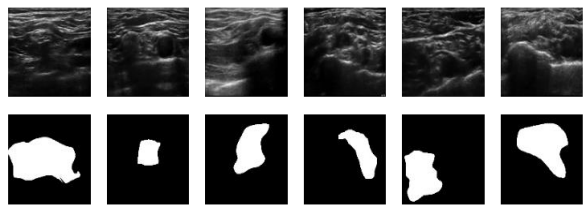

Fig. 1. An example of ultrasound images from each patient (1-6) in this order from left to right and the ground truth: manual annotations (bottom). The white area in the annotated images indicates the annotated nerve segmentation.

TABLE I. ULTRASOUND VIDEOS OF SUPRACLAVICULAR NERVE FROM 6 HEALTHY SUBJECTS ARE CONVERTED TO SUCCESSIVE IMAGES AND ANNOTATED BY TRAINED PHYSICIANS. EACH ANNOTATION IS DONE BY ONE PHYSICIAN. THE TABLE SHOWS NO OF AVAILABLE ANNOTATION FOR EACH SUBJECT.

| SUBJECT NO. | NUMBER OF ANNOTATIONS |
|---|---|
| 1 | 170 |
| 2 | 241 |
| 3 | 265 |
| 4 | 169 |
| 5 | 122 |
| 6 | 151 |

### B. Network Architectures

We modified the original U-Net architecture to include dilated convolution, bilinear interpolation along with skip connections. The U-Net architecture is depicted in Fig. 2. It consists of two pathways: (1) a shrinking pathway which contains series of convolutional layers where input image resolution is reduced, and (2) an expanding pathway which contains Transpose convolution layers to increase the output resolution. The shrinking pathway consists of repeated layers of two 3x3 padded convolutions and a 2x2 max pooling operator with stride 2 for the down-sampling. All the convolutional layers are followed by an advanced rectified linear unit (parametric ReLU [18]). In the expanding pathway, on the other hand, transpose convolution output merge with the tensor from the shrinking pathway whose shape of feature map is the same size as the output, followed by a 3x3 convolution. Since the layers of the expanding pathway have skip connections from the shrinking pathway, the backward propagation of errors is performed without significantly diminishing the gradient value. The tensor after each max pooling layer was used to calculate the binary cross-entropy value corresponding with the true pixel values (labels) of segmentation.

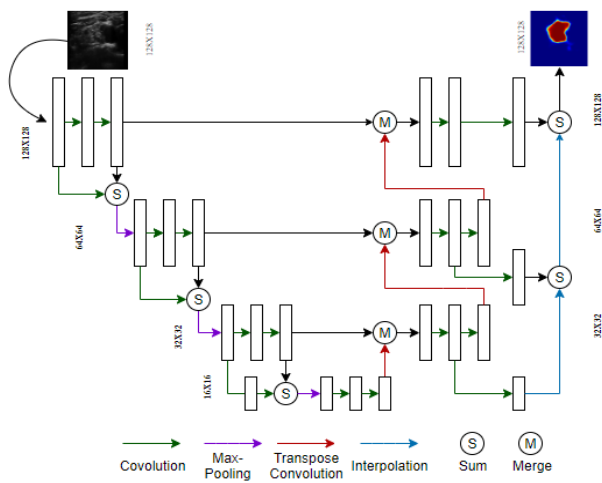

Fig. 2. Modified U-Net Architecture

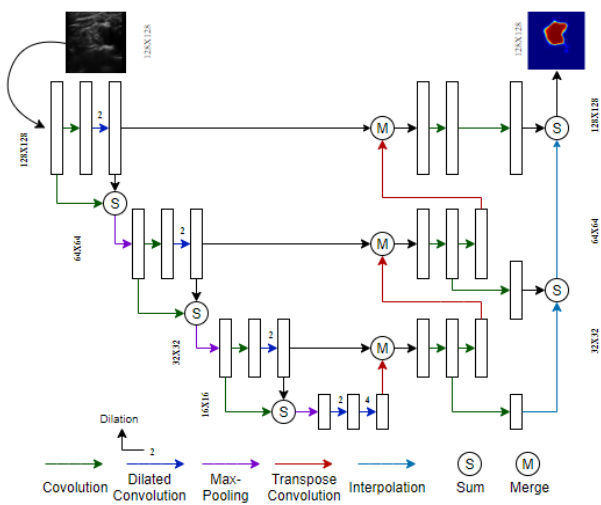

Fig. 3. Dilated U-Net Architecture

Dilated U-Net shown in Fig. 3, uses a similar architecture, with the addition of dilated convolutional layers to achieve increased receptive field in the Dilated U-Net architecture. Dilated convolutions provide the knowledge of the global context of image on how the objects are arranged relative to one another. The neurons in the deepest part of dilated U-Net have larger receptive fields than conventional U-Net which enables the network to see the global context in producing the segmentation mask. At the end of shrinking block of U-Net we have added dilated convolution with increasing dilation starting with dilation 2. These dilated convolution layers are an efficient alternative to increasing the number of layers to achieve increased receptive field. Adding more layers without dilations increases the receptive field at the cost of increasing the network parameters, resulting in overfitting and slow training performance. Dilated convolutions can effectively produce exponentially expanding receptive fields in contrast to linearly increasing conventional convolution layers, while using minimum number of network parameters [12]. The inner most neurons of the dilated U-Net model have receptive field spanning over the entire width and height of the image because of the dilated convolution layers. The convolutions and Transpose convolutions in the expanding branch of U-Net are left unaltered.

*C. Training*

We trained both U-Net and Dilated U-Net models shown in Fig. 2 and Fig. 3 respectively, separately on NVidia Titan X GPU for 40 epochs with early stopping enabled with a patience of 5 epochs. The neural network architectures were implemented using Keras library [19] and Python 3.6. We used Adam optimizer [20] as a gradient descent optimization algorithm, binary cross entropy as a loss function, and validation dice score as the metric to choose best weights for the model. Original ultrasound images were compressed and transformed into the 128x128 pixel images for the input of the networks for the networks. We obtained the segmentation predictions for two different settings: U-Net and Dilated U-Net. These architectures were compared by their dice score, defined as:

$$Dice\ Score\ =\ \frac{2\ *\ |Y_{true}\ \cap\ Y_{pred}|}{|Y_{true}|\ +\ |Y_{pred}|}$$

Where Y_true is true binary pixel values of segmentation and Y_pred is a predicted probability represented as p. All models predict a probabilistic value of each pixel, we used a threshold of 0.5 to convert the prediction to binary value defined as:

$$Y_{pred} = \begin{cases} 1 & if\ p \geq 0.5 \\ 0 & if\ p < 0.5 \end{cases}$$

We used a nested cross validation approach to train the models. Data from six patients were split into training, validation and test sets consisting of four, one and one patients, respectively. We ran the above models for all possible combinations of patients on the train, test and validation sets. Reported metrics in Table II are based on the average of all five runs.

## III. RESULTS

TABLE II: DICE SCORES PRODUCED BY U-NET AND DILATED U-NET NETWORKS

| PATIENT | U-NET | Dilated U-Net |
|---------|-------|---------------|
| Patient 1 | 0.50 | 0.53 |
| Patient 2 | 0.60 | 0.62 |
| Patient 3 | 0.64 | 0.65 |
| Patient 4 | 0.53 | 0.57 |
| Patient 5 | 0.43 | 0.50 |
| Patient 6 | 0.40 | 0.51 |
| Average | 0.52 | 0.56 |

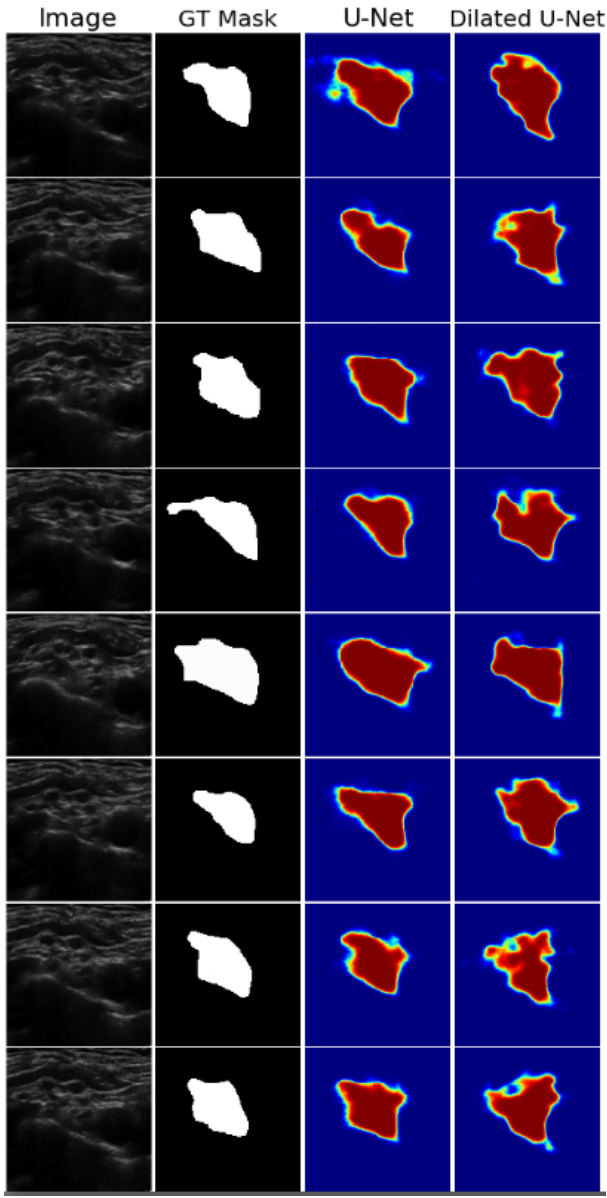

Fig. 4. Successful predictions with dice score (>0.8) predicted by U-Net architecture and the corresponding Dilated U-Net prediction. GT= Ground Truth.

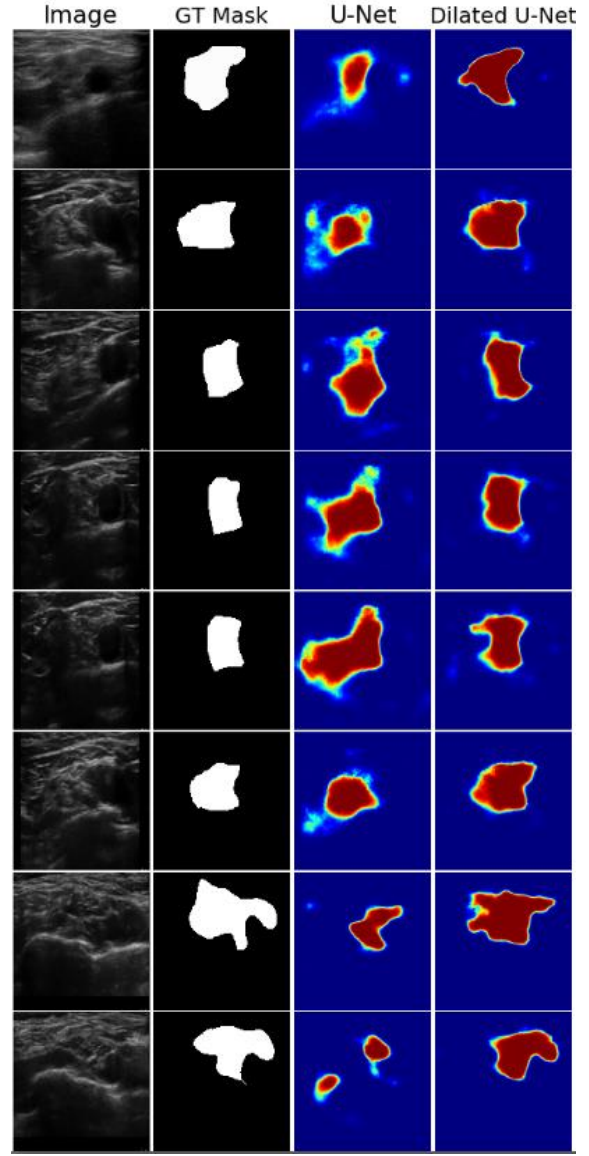

Fig. 5. Successful predictions with dice score (>0.8) predicted by Dilated U-Net architecture and the corresponding U-Net prediction. GT= Ground Truth.

We performed nested cross-validation approach on U-Net and Dilated U-Net models. The Dataset is augmented using random rotations, height, and width shifts. Fig 4 and Fig 5 show predictions of U-Net and dilated U-Net respectively on the same set of images. Fig. 4 corresponds to successful segmentation achieved by U-Net with (dice score > 0.80) and corresponding predictions by Dilated U-Net. Fig. 5 shows successful segmentations by the Dilated U-Net architecture with (dice score > 0.80) and corresponding predictions by U-Net. In Fig. 4 and Fig. 5 we compare images with best predictions from one network with corresponding predictions of same images by the counterpart. It can be observed dilated U-Net managed to achieve dice scores close to the predictions of the U-Net architecture Fig. 4 while U-Net struggled to localize the segmentation and achieve good dice score Fig. 5. Fig.6 and Fig.7 shows sample images where U-Net and Dilated U-Net showed poor performance respectively.

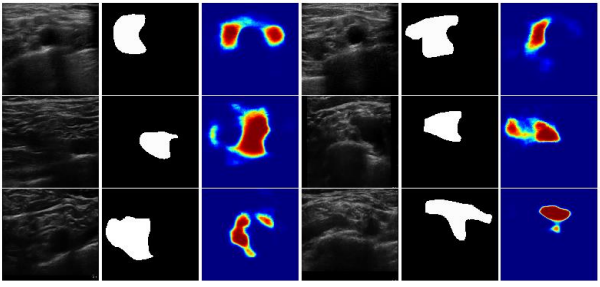

Fig. 6. Example Failed predictions by U-Net. All the predictions shown have dice score (<0.40).

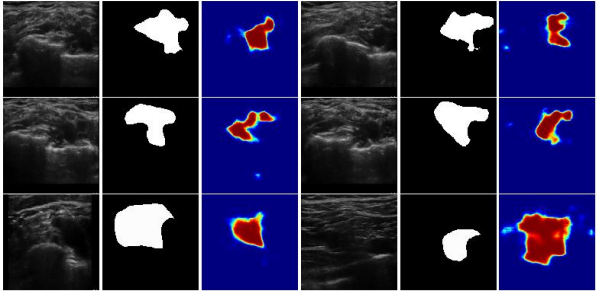

Fig. 7. Example Failed predictions by Dilated U-Net. All the predictions shown have dice score (<0.40).

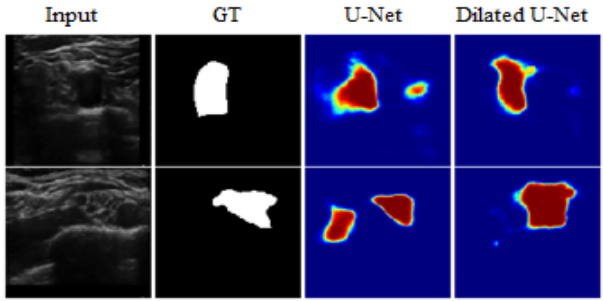

Fig. 8. Comparison of U-Net and Dilated U-Net prediction for random images.

## IV. DISCUSSION

The highest segmentation performance was mostly achieved with dilated U-Net architecture as described in Table II, and it was a dice score of 0.56 average of all the six subjects. Fig. 6 and Fig. 7 suggests even with dice score < 0.40 they correctly identified potential nerve distributions that were atypical in their anatomic orientation around the artery. Despite low dice scores, many of the images suggested clinically reasonable possibilities for local anesthetic deposition.

In Fig. 8, it can be seen U-Net localizes prediction to two different regions one of which is a not a region of interest. The reason for this behavior of the network is due to the limited receptive field of the innermost layers of the network. . None of the neurons in the inner most layers have the context of entire input image. The Network does not realize that there is only one nerve block per image due to the limited receptive field. Receptive field of inner most layers of U-Net is confined to 68 while receptive field dilated U-Net spans across the entire input image.

Our study had several limitations. Ground truth annotations were made based on clinically reasonable regions for local anesthetic deposition. These annotated regions capture not only target nerves but also, the surrounding region's where it is clinically acceptable to inject anesthesia. These manual annotations are not precise in differentiating between nerves of interest and the background. Our dataset is limited to only 6 subjects which is another challenge for the model to learn higher level features and generalize the prediction for varying patient population. These limitations in the manual segmentation for the ground truth and lack large training dataset had affected the model's performance, as shown in Fig. 9.

In the practical use of ultrasound images, doctors are sometimes required to recognize the target object in a timely manner for treatment and diagnosis. Obtained masks could potentially be helpful for trainees. For the same reason, while our methods were unable to definitively determine a comparison of model versus human ability to recognize the true nerve from ultrasound images, our result indicates that our proposed model may be supportive during UGRA applications as well as in developing new research directions in UGRA investigations.

From a machine learning exercise in segmentation, these findings also promoted general knowledge discovery in allowing clinicians to re-review highlighted regions in examples with both high and low dice scores in order to consider a 2nd look examination of regions of interest. Given this, we consider the low dice scores to be at least as interesting from a clinical perspective as those successful segmentations with high dice scores. For perspective, UGRA remains quite novel and dynamic in even basic assumptions of nerve block performance.

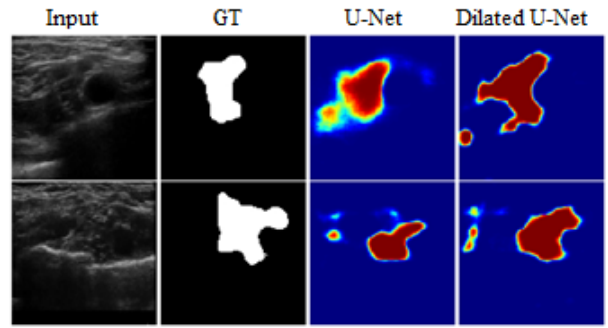

Fig. 9. Comparison of U-Net and dilated U-Net prediction for random images.

We will need to validate these findings with clinical approaches in cases with both successful and failed block and include determinate mapping of involved and spared regions. Such validations may be aided by nerve stimulation studies, although the electrical field may also bias motor versus sensory distributions as well as effect spatial resolution given electrical field size requirements. In addition, future work will be involved in building a model using only poorly performed images. Analysis of feature representation from the model would provide suggestions for further improvements of performance. Also, other future works might be involved in an analysis of our data with temporal models such as a Recurrent Neural Network (RNN) [21-23]. Since our original data were recorded as videos, in our experiments we lost some information by converting video into non-successive images. It is possible to provide successive image to an RNN to utilize the information embedded in sequential frames.

## V. CONCLUSION

We used two different U-Net architectures to predict the Supraclavicular nerve block of six different patients using nested cross validation approach. In one of the models dilated convolution were introduced to effectively increase receptive field of inner most neurons of the network so that the model has context of surrounding anatomical structure. Our results show network with dilated convolutions have shown increased performance than the U-Net architecture. Also dilated convolutions helped the network localize the nerve block to one location which regular U-Net has failed to do so due to limited receptive field of the neurons in the inner most layers. Although the networks did not show clinically acceptable performance in supportive of UGRA applications, our models asserted the ability of deep learning for automatic prediction of nerves in ultrasound images.

## REFERENCES

[1] J. M. Neal *et al.*, "The second American society of regional anesthesia and pain medicine evidence-based medicine assessment of ultrasound-guided regional anesthesia: executive summary," *Regional anesthesia and pain medicine,* vol. 41, no. 2, pp. 181-194, 2016.

[2] S. N. Narouze *et al.*, "The American Society of Regional Anesthesia and Pain Medicine, the European Society of Regional Anaesthesia and Pain Therapy, and the Asian Australasian Federation of Pain Societies Joint Committee recommendations for education and training in ultrasound-guided interventional pain procedures," *Regional anesthesia and pain medicine,* vol. 37, no. 6, pp. 657-664, 2012.

[3] E. Mariano. "Inaugural ACGME-Accredited RAAPM Fellowships Announced." https://www.asra.com/news/147/inaugural-acgme-accredited-raapm-fellows (accessed 2018).

[4] J. M. Neal, "Ultrasound-guided regional anesthesia and patient safety: update of an evidence-based analysis," *Regional anesthesia and pain medicine,* vol. 41, no. 2, pp. 195-204, 2016.

[5] Y. LeCun, Y. Bengio, and G. Hinton, "Deep learning. nature 521 (7553): 436," *Google Scholar,* 2015.

[6] T.-Y. Lin *et al.*, "Microsoft coco: Common objects in context," in *European conference on computer vision*, 2014: Springer, pp. 740-755.

[7] A. Krizhevsky, I. Sutskever, and G. E. Hinton, "Imagenet classification with deep convolutional neural networks," in *Advances in neural information processing systems*, 2012, pp. 1097-1105.

[8] J. Deng, W. Dong, R. Socher, L.-J. Li, K. Li, and L. Fei-Fei, "Imagenet: A large-scale hierarchical image database," in *Computer Vision and Pattern Recognition, 2009. CVPR 2009. IEEE Conference on*, 2009: Ieee, pp. 248-255.

[9] M. Everingham, L. Van Gool, C. K. Williams, J. Winn, and A. Zisserman, "The pascal visual object classes (voc) challenge," *International journal of computer vision,* vol. 88, no. 2, pp. 303-338, 2010.

[10] J. Xiao, J. Hays, K. A. Ehinger, A. Oliva, and A. Torralba, "Sun database: Large-scale scene recognition from abbey to zoo," in *Computer vision and pattern recognition (CVPR), 2010 IEEE conference on*, 2010: IEEE, pp. 3485-3492.

[11] O. Ronneberger, P. Fischer, and T. Brox, "U-net: Convolutional networks for biomedical image segmentation," in *International Conference on Medical image computing and computer-assisted intervention*, 2015: Springer, pp. 234-241.

[12] F. Yu and V. Koltun, "Multi-scale context aggregation by dilated convolutions," *arXiv preprint arXiv:1511.07122,* 2015.

[13] M. Drozdzal, E. Vorontsov, G. Chartrand, S. Kadoury, and C. Pal, "The importance of skip connections in biomedical image segmentation," in *Deep Learning and Data Labeling for Medical Applications*: Springer, 2016, pp. 179-187.

[14] C. Liu, F. Liu, L. Wang, L. Ma, and Z.-M. Lu, "SEGMENTATION OF NERVE ON ULTRASOUND IMAGES USING DEEP ADVERSARIAL NETWORK," *INTERNATIONAL JOURNAL OF INNOVATIVE COMPUTING INFORMATION AND CONTROL,* vol. 14, no. 1, pp. 53-64, 2018.

[15] K. He, X. Zhang, S. Ren, and J. Sun, "Deep residual learning for image recognition," in *Proceedings of the IEEE conference on computer vision and pattern recognition*, 2016, pp. 770-778.

[16] C. A. Schneider, W. S. Rasband, and K. W. Eliceiri, "NIH Image to ImageJ: 25 years of image analysis," *Nature methods,* vol. 9, no. 7, p. 671, 2012.

[17] V. P. S. Fazan, A. d. S. Amadeu, A. L. Caleffi, and O. A. Rodrigues Filho, "Brachial plexus variations in its formation and main branches," *Acta Cirurgica Brasileira,* vol. 18, pp. 14-18, 2003.

[18] K. He, X. Zhang, S. Ren, and J. Sun, "Delving deep into rectifiers: Surpassing human-level performance on imagenet classification," in *Proceedings of the IEEE international conference on computer vision*, 2015, pp. 1026-1034.

[19] F. Chollet, "Keras (2015)," ed, 2017.

[20] D. P. Kingma and J. Ba, "Adam: A method for stochastic optimization," *arXiv preprint arXiv:1412.6980,* 2014.

[21] S. Hochreiter and J. Schmidhuber, "Long short-term memory," *Neural computation,* vol. 9, no. 8, pp. 1735-1780, 1997.

[22] F. A. Gers, J. Schmidhuber, and F. Cummins, "Learning to forget: Continual prediction with LSTM," 1999.

[23] J. Donahue *et al.*, "Long-term recurrent convolutional networks for visual recognition and

description," in *Proceedings of the IEEE conference on computer vision and pattern recognition*, 2015, pp. 2625-2634.